\documentclass[twocolumn,showpacs,preprintnumbers,amsmath,amssymb]{revtex4}

\usepackage{amsmath,amssymb,graphicx}
\usepackage{graphicx}
\usepackage{dcolumn}
\usepackage{bm}
\usepackage{epsfig}
\usepackage{float}
\usepackage{amsmath}
\usepackage[usenames]{color}

\newcommand{\bea}{\begin{eqnarray}}
\newcommand{\eea}{\end{eqnarray}}

\begin{document}
\draft


\title{Analysis of recent type Ia supernova data based
       on evolving dark energy models}
\author{Jaehong Park, Chan-Gyung Park and Jai-chan Hwang}
\address{Department of Astronomy and Atmospheric Sciences,
                 Kyungpook National University, Daegu, Korea}

\date{\today}

\begin{abstract}
We study characters of recent type Ia supernova (SNIa) data using
evolving dark energy models with changing equation of state
parameter $w$. We consider sudden-jump approximation of $w$ for some
chosen redshift spans with double transitions, and constrain these
models based on Markov Chain Monte Carlo (MCMC) method using the
SNIa data (Constitution, Union, Union2) together with baryon
acoustic oscillation $A$ parameter and cosmic microwave background
shift parameter in a flat background. In the double-transition model
the Constitution data shows deviation outside $1 \sigma$ from
$\Lambda\textrm{CDM}$ model at low ($z \lesssim 0.2$)
and middle ($0.2 \lesssim z \lesssim 0.4$) redshift bins
whereas no such deviations are noticeable in the Union and Union2 data.
By analyzing the Union members in the Constitution set, however, we show that the
same difference is actually due to
different calibration of the same Union sample in the Constitution set,
and is \textit{not} due to new data added in the
Constitution set. All detected deviations are within $2 \sigma$ from
the $\Lambda$CDM world model.
From the $\Lambda\textrm{CDM}$ mock data analysis, we quantify
{\it biases} in the dark energy equation of state parameters induced
by insufficient data with inhomogeneous distribution of data points
in the redshift space and distance modulus errors.
We demonstrate that location of peak in the distribution of arithmetic
means (computed from the MCMC chain for each mock data) behaves as
an unbiased estimator for the average bias, which is valid even
for non-symmetric likelihood distributions.
\end{abstract}

\noindent 

\maketitle

%
%
\section{Introduction}
\label{sec:introduction}

The type Ia supernova (SNIa) observation indicates current accelerated
expansion of the universe \cite{Riess-etal-1998,Perlmutter-etal-1999}.
In order to explain the acceleration the concept of dark energy was
introduced with many different realizations using a fluid, a field, etc.
The SNIa data are important to study possible evolution of dark energy,
often parametrized by an equation of state parameter $w=p/\rho c^2$;
$p$ is the pressure and $\rho c^2$ is the energy density of the dark
energy component. Although the SNIa data have been continuously improving
\cite{Davis-etal-2007,Astier-etal-2005,Riess-etal-2006,Kowalski-etal-2008,
Hicken-etal-2009,Lampeitl-etal-2009}, it is still remarkable that
all current observational data are well accommodated by the simple
$\Lambda$CDM model with $w = -1$. Thus, at the moment the observational
data do not seriously demand evolution in the dark energy equation of state.

Although $\Lambda$CDM model explains observational data well,
diverse time-varying dark energy models and analysis methods have been
introduced to deal with increasing amount of data
\cite{Efstathiou-1999,Huterer-etal-2000,Weller-etal-2000,Chevallier-etal-2000,
Linder-2002,Huterer-etal-2002,Wang-2009,Corasaniti-etal-2002,Alam-etal-2003,
Jassal-etal-2004,Daly-etal_2003,Daly-etal-2004,Wetterich-2004,Gong-2004,
Gong-etal-2005,Kim-etal-2003,dePutter-etal-2007,Shafieloo-etal-2006,
Shafieloo-2007,Hojjati-etal-2009,Jimenez-etal-2009,Linder-etal-2005,
Huterer-etal-2004,Sullivan-etal-2007,Shafieloo-etal-2009,Serra-etal-2009,Qi-etal-2009,
Huang-etal-2009,Gong-etal-2009,Cai-etal-2010,Zhao-etal-2009,Wang-etal-2010}.
Possible evolution of dark energy has been probed mainly by
Chevallier-Polarski-Linder (CPL) parametrization $w(a)=w_0+w_a(1-a/a_0)$
\cite{Chevallier-etal-2000,Linder-2002}
and the redshift-binned parametrization method based on the decorrelating
technique \cite{Huterer-etal-2002,Huterer-etal-2004,Sullivan-etal-2007}.
In the latter, the equation of state parameter in each redshift bin is
estimated independently of others based the piecewise constant, linear
or cubic spline parametrizations. Compared with the CPL parametrization,
which has only two free parameters with special evolution,
the redshift-binning method is more suitable to know specific dark energy
variations because we can divide the redshift interval into many bins as we
want. Although there is a restriction on the number of redshift bins due to
insufficient data, the redshift-binning method is more appropriate to study
the nature of dark energy.

There are several works showing tendency of $w$ deviating from $-1$
especially in the low redshift region
\cite{Shafieloo-etal-2009,Serra-etal-2009,Huang-etal-2009,Gong-etal-2009,
Cai-etal-2010,Zhao-etal-2009,Wang-etal-2010}. The tendency indicates slow down of
acceleration with the deceleration parameter $q$ increased in $z
\leq 0.3$ \cite{Shafieloo-etal-2009}. Similarly, authors of Ref.\
\cite{Huang-etal-2009} showed that $w$ is rapidly decreased as $z$
increases under both the CPL and the redshift-binned parametrization
methods. In Ref.\ \cite{Cai-etal-2010}, the linear spline method was
used to show that $w(z)$ can be oscillating around $w=-1$ with two
turning points where $w < -1$ at $z\simeq 0.4$ and $w > -1$ at
$z\simeq 1$. The authors of Ref.\ \cite{Zhao-etal-2009} also
reported weak hint of possible evolution of dark energy; it is found
that under the piecewise constant parametrization $w$ is less than
$-1$ in $z$ between $0.25$ and $0.75$. On the other hand, no
significant evidence for dark energy evolution is found based on the
similar methods in Refs.\ \cite{Gong-etal-2009, Serra-etal-2009,Wang-etal-2010}.
In the high redshift region, authors of Ref.\ \cite{Qi-etal-2009} report
that the dark energy deviates from the cosmological constant at redshift
$z \gtrsim 0.5$ based on analysis of SNIa and gamma-ray bursts (see also
Ref.\ \cite{Cai-etal-2010}).

In this paper we study the evolution of dark energy using a double
sudden-jump transition model with varying transition redshift, which
is a simple version of the piecewise constant parametrization
method. To constrain our model we use the recent SNIa data such as
Union \cite{Kowalski-etal-2008}, Constitution
\cite{Hicken-etal-2009}, Union2 \cite{Amanullah-etal-2010} samples,
together with the baryon acoustic oscillation (BAO) parameter
\cite{Eisenstein-etal-2005} and the cosmic microwave background
(CMB) shift parameter \cite{Komatsu-etal-2008,Bond-etal-1997,
Wang-etal-2006}. Especially we investigate the characters of the recent
SNIa data sets in view of the simple evolving dark energy models
($i$) by comparing the results of the Union, the Constitution data
sets, and the same Union members included in the Constitution sample
and ($ii$) by analyzing the $\Lambda\textrm{CDM}$ based mock data
sets mimicking the Constitution and Union2 samples.

According to our double-transition model, we found that the Union sample
and the same Union members in the Constitution sample,
both of which are based on the same light-curve fit parameters,
show different behaviors of dark energy equation of state due to
different calibration processes applied to the distance moduli of the
Union members.
From the $\Lambda\textrm{CDM}$ mock data analysis, we also found that
inhomogeneous distribution of SNIa data points in the redshift space
and inhomogeneous distance modulus errors cause biases in the dark energy
equation of state parameters. We have demonstrated that the peak of the distribution
of arithmetic means (estimated from thousands of mock data sets) can be
used as the unbiased estimator for the average bias in the dark energy
equation of state parameters.

The paper is organized as follows. We introduce our dark energy models
in Sec.\ \ref{sec:model} and describe our analysis method and the
observational data used in this study in Sec.\ \ref{sec:method-data}.
Our results are presented in Secs.\ \ref{sec:double} and \ref{sec:charac}.
Section \ref{sec:discussion} is a discussion of our analysis with conclusions.

%
%
\section{Dark energy model} \label{sec:model}

We consider a $w\textrm{CDM}$ model dominated by dark energy fluid
with equation of state parameter $w$ and cold dark matter. For
generally varying dark energy equation of state the Friedmann
equation is given by
\begin{widetext}
\bea
   & & E^2(z) \equiv \frac{H^2(z)}{H_{0}^{2}}
       = \Omega_{\textrm{r}0}(1+z)^4  + \Omega_{\textrm{m}0}(1+z)^3
          + \Omega_{\textrm{DE}0}f(z) + \Omega_{K0}(1+z)^2,
   \label{eq:wCDM-Friedmann} \\
   & & f(z) \equiv e^{3\int_{0}^{z} [1+w(z)] d \ln{(1+z)}},
   \label{eq:fz}
\eea
\end{widetext}
where $w(z)$ is the dark energy equation of state parameter at
redshift $z$; $z \equiv a_0/a-1$ with $a(t)$ the cosmic expansion
scale factor; $H\equiv {\dot a}/a$ is the Hubble parameter; a dot
represents a derivative with respect to the cosmic time $t$;
$\Omega_{\textrm{r}}$, $\Omega_{\textrm{m}}$, and
$\Omega_{\textrm{DE}}$ are radiation, matter, and dark energy
density parameters, respectively; $\Omega_{K}=-K/(a^2 H^2)$;
subscript $0$ indicates present epoch. We set the present CMB
temperature $T_0=2.725~\textrm{K}$ and the number of massless
neutrino species $N_{\nu}=3.04$. In this work we consider dark
energy models with piecewise constant dark energy equation of state,
in which $w$ parameter is constant for given redshift bins with a
sudden jump at transitions. As a simple case, we consider a
double-transition model (with three redshift bins).  By imposing
conditions that $a$ and $\dot a$ are continuous at the transition
epochs, we rewrite Eq.\ (\ref{eq:fz}) as
\begin{widetext}
\bea
    & & f(z) = \left\{ \begin{array}{lc}
          (1+z)^{3(1+w_{0})}
          & \quad z < z_{\textrm{tr}1} \\
          (1+z)^{3(1+w_{1})}(1+z_{\textrm{tr}1})^{3(w_0-w_1)}
          & \quad z_{\textrm{tr}1} \leq z < z_{\textrm{tr}2} \\
          (1+z)^{3(1+w_{2})}(1+z_{\textrm{tr}1})^{3(w_0-w_1)}(1+z_{\textrm{tr}2})^{3(w_1-w_2)}
          & \quad z \geq z_{\textrm{tr}2,}
          \end{array} \right.
    \label{eq:2trz-friedmann}
\eea
\end{widetext}
where $w_{i}$ ($i=0,1,2$) is the dark energy equation of state parameter
in each redshift bin; $z_{\textrm{tr}1}$ and $z_{\textrm{tr}2}$ represent
the locations of transition. The general formula with arbitrary number of bins
can be found in Refs.\ \cite{Sullivan-etal-2007,Huang-etal-2009}.
We assume a flat background $\Omega_{K}=0$, thus
$\Omega_{\textrm{DE}}=1-\Omega_\textrm{r}-\Omega_\textrm{m}$. As we will consider models with
fixed transition redshifts throughout this paper, we have five free parameters
for our double-transition model. We denote the parameters as
$\mbox{\boldmath $\theta$} =(h,\Omega_{\textrm{m}0},w_0, w_1, w_2)$;
$h$ is a normalized present day Hubble parameter
$H_0 \equiv 100h~\textrm{km}~\textrm{s}^{-1}~\textrm{Mpc}^{-1}$.

%
%
\section{Method and data}
\label{sec:method-data}

\subsection{Markov Chain Monte Carlo method}
\label{subsec:mcmc}

To obtain likelihood distributions for parameters $\mbox{\boldmath
$\theta$}$, we use the Markov Chain Monte Carlo (MCMC) method.
The MCMC chain elements are generated to randomly explore the whole
parameter space based on the Metropolis-Hastings algorithm
\cite{Metropolis-etal-1953,Hastings-1970}. For the probability density
function $P(\mbox{\boldmath $\theta$}|{\mathbf D})$ which is needed to make
decisions for accepting/rejecting a randomly chosen chain element, we use
\begin{equation}
    P(\mbox{\boldmath $\theta$} | {\mathbf D})
        \propto \exp \left(-\frac{\chi^2}{2} \right), \quad
        \chi^{2}=\sum_{i=1}^{N}\frac{[X_{i}(\mbox{\boldmath $\theta$})
        -X_{\textrm{obs},i}]^{2}}{\sigma_{\textrm{obs},i}^{2}},
    \label{eq:chi-square}
\end{equation}
where $\mathbf D$ denotes data, $X_{i}(\mbox{\boldmath $\theta$})$ represents
the model prediction for the $i$th observed data point $X_{\textrm{obs},i}$
with measurement error $\sigma_{\textrm{obs},i}$, and $N$ is the total number
of data points. We use SNIa, BAO, and CMB data to constrain our model. In
this case the final $\chi^2$ is the sum of individual $\chi^2$'s:
$\chi^2=\chi_\textrm{SN}^2+\chi_\textrm{BAO}^2+
\chi_\textrm{CMB}^2$.

In applying the MCMC method to our dark energy models,
we impose two priors.
First, we impose a prior that the dark energy at sufficiently early epoch
($z_i=10^{8}$) should be lower than the maximum amount allowed by
the big bang nucleosynthesis calculation, $\Omega_{\textrm{DE}i} < 0.045$
\cite{BBN}.
If the number and the quality of SNIa data points within a redshift bin
are not sufficient, the MCMC chain of the dark energy equation of state
parameter of the bin usually does not converge, gradually decreasing to smaller
and smaller values, sometimes reaching $-100$ in an extreme case.
Therefore, as the second prior we set lower bounds to all the dark energy
equation of state parameters, $w_i > -5$, to make sure the convergence
of the MCMC chain.

In applying the MCMC method to the observational data, we generate
one million chain elements after a burning period during which the
first two thousand chain elements are discarded. For a
$\Lambda\textrm{CDM}$ based mock data set (see Sec.\
\ref{subsec:mockdata} below), we generate $0.2$ million chain
elements. To test the convergence of MCMC chains we use a simple
diagnostic: the means calculated from the first (after burning
process) and the last 10\% of the chain becomes approximately equal
to each other if the chain has converged (see Appendix B of Ref.\
\cite{Abrahamse-etal-2008} for detailed description). We found that
the $0.2$ million chain elements are sufficient for the convergence
of our MCMC chains. For each dark energy equation of state parameter
in our models, we obtain its one-dimensional likelihood distribution
by marginalizing over all other parameters, and compute the
one-dimensional likelihood peak value or the arithmetic mean value
as our average estimate, together with $1\sigma$ ($68.3$\%) and
$2\sigma$ ($95.4$\%) confidence limits.

It should be noted that in our simple piecewise constant parametrization
of dark energy equation of state we do not apply the decorrelating technique
to obtain the independent estimate of $w_i$'s.
Instead, by comparing the $w_i$'s measured from the observational
data with the likelihood confidence regions that the $\Lambda\textrm{CDM}$
mock data analysis allows, we obtain a desirable statistical criterion
for whether the measured deviation can be interpreted as dark energy
evolution or not.

\subsection{Observational data}
\label{subsec:obsdata}

To constrain our dark energy model we jointly use the three kinds of
observational data such as the SNIa data, BAO $A$, and CMB $R$
parameters.

For the SNIa data, we use the Union sample (307 members)
\cite{Kowalski-etal-2008}, the Constitution sample (397 members)
\cite{Hicken-etal-2009}, and the Union2 sample (557 members)
\cite{Amanullah-etal-2010}.
For Union and Union2 data analysis, we use the covariance matrix
between distance modulus errors without systematics.
The Constitution data set consists of the same SNIa members in the Union
data set and 90 CfA3 SNIa added at low redshift ($z<0.1$).
We will analyze four data sets: the Constitution set, the Union set,
pure Union sample present in the Constitution set (hereafter Constitution-U),
and Union2 set.
The SNIa distance modulus in the Constitution-U set is different from
the one in the original Union set due to a new calibration used in the
Constitution data set.

\begin{figure}[t]
\begin{center}
\includegraphics[width=78mm]{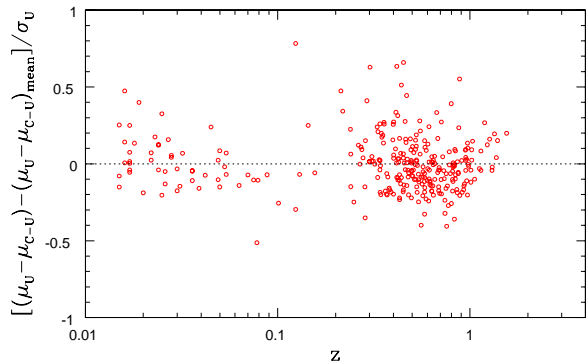}
\caption{
         A plot showing the difference between Constitution-U
         ($\mu_\textrm{C-U}$) and Union ($\mu_\textrm{U}$) SNIa
         distance modulus relative to the measurement error.
         Since each data set has an arbitrary zero point in the distance
         modulus,
         actually shown are mean-subtracted differences
         between two distance moduli divided by the distance
         modulus uncertainty of the Union SNIa ($\sigma_\textrm{U}$).
         }
\label{fig:Union-CU}
\end{center}
\end{figure}

Authors of Ref.\ \cite{Hicken-etal-2009} produced the Constitution data set
by taking directly the SALT light-curve fit parameters of the Union set
(listed in Table 11 of Ref.\ \cite{Kowalski-etal-2008}) and
adding them to the SALT output of CfA3 sample.
They found the best-fit cosmology from the combined
SALT output parameters (rest-frame peak magnitude $m_B$, stretch $s$,
color $c$) by using the empirical relation for the distance modulus
\begin{equation}
   \mu = m_B - M_B + \alpha (s-1) - \beta c,
\end{equation}
obtaining empirical coefficients related to stretch correction
($\alpha=1.34_{-0.08}^{+0.08}$), color correction
($\beta=2.59_{-0.08}^{+0.12}$), and absolute magnitude ($M_B$),
which are different from those used in the Union data analysis
($\alpha=1.24\pm 0.10$, $\beta=2.28\pm 0.11$; \cite{Kowalski-etal-2008}).
Furthermore, in the Constitution data set,
a host galaxy peculiar velocity of $400~\textrm{km}~\textrm{s}^{-1}$
has been assumed (rather than $300~\textrm{km}~\textrm{s}^{-1}$ used
in the Union data set \cite{Kowalski-etal-2008}),
and the distance modulus uncertainties of Union SNIa members have been
modified to reproduce the same amount of uncertainty in the measurement
of $w$ (based on the constant $w$ dark energy model) as in Ref.\
\cite{Kowalski-etal-2008}.
Besides, an uncertainty of $0.138$ mag has been added in quadrature to
the uncertainties of SALT light-curve fit for each CfA3 SNIa.
Since different empirical coefficients and different level of additional
uncertainties were used to calculate the distance modulus of each SNIa,
the Union and the Constitution sets have experienced different calibration
process (see Ref.\ \cite{Hicken-etal-2009} for a complete description).
Figure \ref{fig:Union-CU} compares distance moduli of Constitution-U
and Union SNIa samples, where for each SNIa member the mean-subtracted
difference between two distance moduli divided by the measurement uncertainty
is shown. Although distance modulus values from two samples
do not show special trend, the plot shows a scatter in both positive and negative
directions, implying that SNIa members of Constitution-U and Union samples
experienced the different calibration process.

For a given background world model the distance modulus at redshift
$z$ is given by
\begin{equation}
    \mu \equiv m-M=5\log_{10}\left[ \frac{(1+z)r(z)}{1~\textrm{Mpc}}\right]+25,
    \label{eq:distance-modulus}
\end{equation}
with the comoving distance at redshift $z$ given as
\begin{equation}
    r(z)=\frac{c}{H_0\sqrt{\Omega_{K0}}}
       \sin \left[ \sqrt{\Omega_{K0}} \int^{z}_{0} \frac{H_0}{H(z)} dz
       \right].
\label{eq:comoving-distance}
\end{equation}
The SNIa data provide a distance modulus $\mu_{\textrm{obs},i}$
with a measurement error $\sigma_{\textrm{obs},i}$ for each SNIa
at redshift $z_i$. We obtain $\chi^2$ for the SNIa data as
\begin{equation}
    \chi_{\textrm{SN}}^{2}
       =\sum_{i=1}^{N_{\rm SN}}\frac{[\mu_{i}(\mbox{\boldmath $\theta$})
        -\mu_{\textrm{obs},i}]^{2}}{\sigma_{\textrm{obs},i}^{2}},
    \label{eq:chi2-SN}
\end{equation}
where $N_{\rm SN}$ is the number of SNIa in the data set, and
$\mu_i(\mbox{\boldmath $\theta$})$ is the distance modulus at
redshift $z_i$ predicted by the theoretical world model to be
constrained.


The BAO $A$ parameter is defined as \cite{Eisenstein-etal-2005}
\begin{equation}
    A\equiv\Omega_{\textrm{m}0}^{1/2}E(z_{b})^{-1/3}
         \left[ \frac{1}{z_b}\int_{0}^{z_{b}}\frac{dz'}{E(z')}
         \right]^{2/3}.
    \label{eq:A-parameter-the}
\end{equation}
From the Sloan Digital Sky Survey luminous red galaxy sample we have
\cite{Eisenstein-etal-2005}
\begin{equation}
    z_b = 0.35, \quad
    A_{\textrm{obs}}=0.469\left(\frac{n_{s}}{0.98}\right)^{-0.35} \pm 0.017,
    \label{eq:A-parameter-obs}
\end{equation}
where $n_s$ is the spectral index of scalar-type perturbation. We take
$n_s = 0.960$ based on the Wilkinson Microwave Anisotropy Probe (WMAP)
5-year results \cite{Komatsu-etal-2008}. The $\chi^2$ of BAO data is
\begin{equation}
    \chi_{\textrm{BAO}}^{2}=\frac{[A(\mbox{\boldmath $\theta$})
                -A_{\textrm{obs}}]^{2}}{\sigma_{\textrm{obs},A}^{2}},
    \label{eq:chi2-BAO}
\end{equation}
with $\sigma_{\textrm{obs},A}=0.017$ \cite{Eisenstein-etal-2005}.


We use CMB shift parameter $R$ introduced as
\cite{Bond-etal-1997,Wang-etal-2006}
\begin{equation}
   R\equiv\ \sqrt{\Omega_{\textrm{m}0}H_{0}^{2}/c^2}\cdot r(z_{\ast}),
   \label{eq:R-parameter-the}
\end{equation}
where $z_{\ast}$ is the redshift of recombination. From the WMAP
5-year data we have \cite{Komatsu-etal-2008}
\begin{equation}
    z_{\ast}=1090, \quad
    R_{\textrm{obs}}=1.710 \pm 0.019.
    \label{eq:R-parameter-obs}
\end{equation}
The $\chi^2$ of the $R$ parameter is
\begin{equation}
    \chi_{\textrm{CMB}}^{2}=\frac{[R(\mbox{\boldmath $\theta$})
          -R_{\textrm{obs}}]^{2}}{\sigma_{\textrm{obs},R}^{2}},
    \label{eq:chi2-CMB}
\end{equation}
with $\sigma_{\textrm{obs},R}=0.019$ \cite{Komatsu-etal-2008}.

\subsection{Mock data}
\label{subsec:mockdata}

In our estimation of $w_i$ at each redshift bin, the deviation from
$\Lambda$CDM value can be interpreted as the evolution of dark
energy. However, even the observational data obtained in a perfect
$\Lambda\textrm{CDM}$ world may give a spurious deviation from
$w=-1$ due to inhomogeneous distribution of data points in the
redshift space and the measurement errors, which makes the detection
of dark energy evolution more complicated. To include such a
spurious statistical effect in our analysis, we have generated
thousands of $\Lambda\textrm{CDM}$-motivated mock data sets and
analyzed them in a similar way as the real observational data is
analyzed.

For SNIa mock data, we make the five thousand Constitution and Union2
mock data sets predicted in the $\Lambda\textrm{CDM}$ model by using
redshifts ($z_i$'s) and distance modulus measurement errors
($\sigma_{\textrm{obs},i}$'s) of SNIa members in the Constitution
and the Union2 samples. For $\Lambda\textrm{CDM}$ background world
model we use cosmological parameters that are consistent with the WMAP
5-year observation, i.e.,
$h=0.705$, $\Omega_{\textrm{m}0}=0.274$, $\Omega_{K0}=0$, and $w=-1$
\cite{Komatsu-etal-2008}. For a given SNIa at redshift $z_i$, the mock
data point is generated by
$\mu_{\textrm{mock},i}=\mu_{\Lambda\textrm{CDM},i}+e_i$,
where $\mu_{\Lambda\textrm{CDM},i}$ is a distance modulus predicted
in the $\Lambda\textrm{CDM}$ model and $e_i$ is a random number drawn
from a Gaussian normal distribution with variance $\sigma_{\textrm{obs},i}^2$,
which mimics the measurement noise.
In this manner, we can get $\Lambda\textrm{CDM}$ based SNIa mock data sets
with the same redshift distribution and distance modulus precision of
Constitution and Union2 samples. The same method is applied to generating
mock BAO $A$ and CMB $R$ parameters.
The similar technique of generating SNIa mock data sets has been used in
Refs. \cite{Antoniou-2010, Holsclaw-etal-2010}.

There are two good points of using $\Lambda\textrm{CDM}$ mock data sets.
First, we can include the statistical effect induced by different random
realizations of measurement errors, which usually appears as additional
variance in our estimation of dark energy equation of state parameters.
Thus, the confidence regions of $w_i$'s obtained from the $\Lambda\textrm{CDM}$
mock data sets become bigger than those from the single data set (see Figs.\
\ref{fig:2trz-constitution-mock} and \ref{fig:2trz-union2-mock} below).
By comparing results from the observational data set
with those from $\Lambda\textrm{CDM}$ mock data sets, we can avoid a false
signature of dark energy evolution due to statistically peculiar measurement
noise in the observational data.
Analyzing the single data set cannot account for such a statistical effect.
Second, we can estimate any bias in the estimate of $w_i$ relative to $-1$,
induced by the sparse number of data points and measurement errors within
the redshift bin considered. We investigate this issue in detail
in Sec.\ \ref{sec:charac}.


\section{Constraints on dark energy equation of state
in double-transition models}
\label{sec:double}

\begin{figure*}
\begin{center}
\includegraphics[width=14cm]{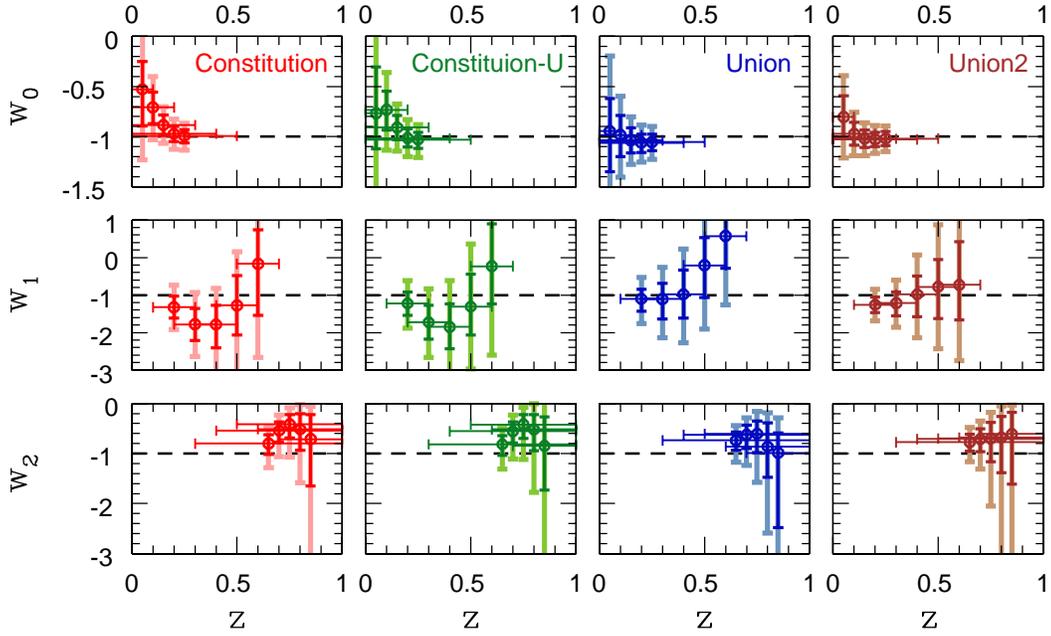}
\caption{Estimated $w_0$, $w_1$, and $w_2$ in the double-transition dark
         energy models. The transition redshift $z_{\textrm{tr}1}$ is varied
         from $0.1$ to $0.5$ by $0.1$ increase in redshift while the width
         between the two transitions is fixed as
         $z_{\textrm{tr}2}-z_{\textrm{tr}1}=0.2$.
         Red, green, blue, and brown colors represent Constitution,
         Constitution-U, Union, and Union2, respectively. The error bars
         with dark (light) color indicate $68.3$\% ($94.5$\%) confidence
         limits.}
\label{fig:2trz-data-plot}
\end{center}
\end{figure*}

In this section, we present the results of our dark energy model
with double transitions. Figure \ref{fig:2trz-data-plot} shows
$w_0$, $w_1$, and $w_2$ measured from Constitution, Constitution-U,
Union, and Union2 SNIa samples together with BAO and CMB data, for
various transition redshifts. We fix the width between the two
transitions as $z_{\rm tr2}-z_{\rm tr1}=0.2$, and vary $z_{\rm tr1}$
from 0.1 to 0.5 with steps of $0.1$. From the marginalized
one-dimensional likelihood distributions of $w_0$, $w_1$, and $w_2$
parameters, we have estimated peak locations (denoted as P) and
$1\sigma$ ($68.3$\%), $2\sigma$ ($95.4$\%) confidence limits (points
with error bars). In the Constitution and Constitution-U sets $w_0$
is larger than $-1$ and $w_1$ is smaller than $-1$ for
$z_{\textrm{tr}1}\le 0.4$. However, such trends are not seen in the
Union and Union2 sets. Values of $w_2$ tend to be larger than $-1$
in all four data sets. Since the Constitution-U sample is composed
of exactly the same SNIa members of the Union data set, we can
conclude that such different behaviors of $w_0$ and $w_1$ arise not
due to the new CfA3 data added in the Constitution set but due to
the different calibration applied to each data set.

\begin{figure*}[htbp]
\begin{center}
\includegraphics[width=14cm]{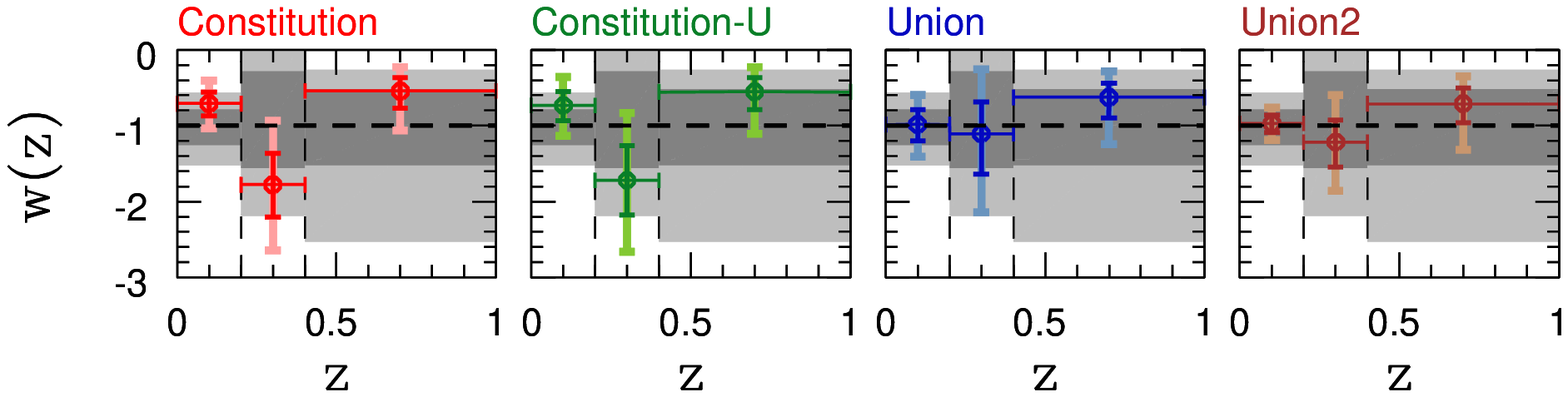}
\includegraphics[width=14cm]{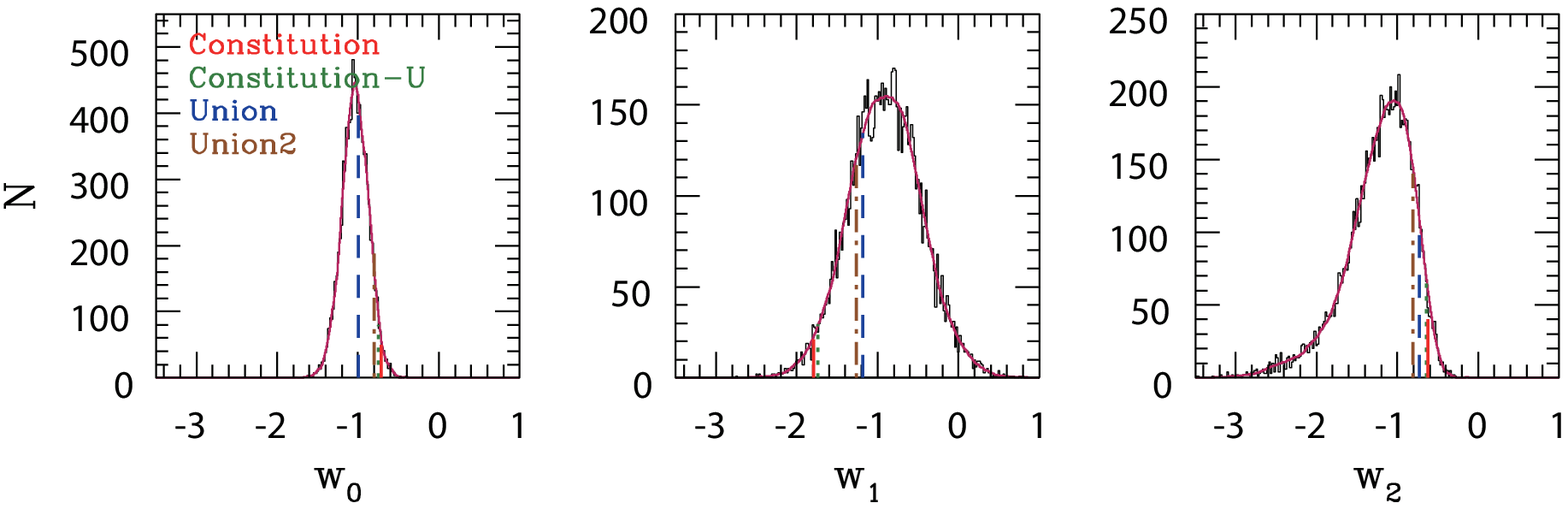}
\caption{
         Top panels:
         Estimated $w_0$, $w_1$, and $w_2$ in the double-transition
         dark energy model with $z_{\textrm{tr}1}=0.2$ and
         $z_{\textrm{tr}2}=0.4$, taken from Fig.\ \ref{fig:2trz-data-plot}
         with the same colored code.
         Data points with error bars are based
         on the actual SNIa data sets.
         In all panels, shaded regions indicating $68.3$\% (dark grey)
         and $95.4$\%
         (light grey) confidence limits have been obtained from the five
         thousand $\Lambda$CDM based Constitution mock data sets.
         Bottom panels:
         Distributions of arithmetic mean (M) values of $w_0$, $w_1$,
         and $w_2$ obtained from the Constitution mock data sets (histograms
         fitted by dark red smooth curves, see text).
         The three bins in the Constitution set have $147$, $49$, and $201$ SNIa
         data points, respectively.
         Red (solid), green (dotted), blue (dashed), and brown (dot-dashed) vertical lines denote locations
         of M values for Constitution, Constitution-U, Union, and Union2 data
         sets, respectively.
         }
\label{fig:2trz-constitution-mock}
\end{center}
\end{figure*}

Now, we consider a case of $z_{\textrm{tr}1}=0.2$ and $z_{\textrm{tr}2}=0.4$
where the deviations in Constitution and Constitution-U sets are
most significant. The result is shown in the top row of Fig.\
\ref{fig:2trz-constitution-mock}. In the Constitution and
Constitution-U sets the $1 \sigma$ confidence limits of {\it all}
$w_i$'s deviate from $w=-1$. However, in the cases of Union and
Union2, $w_0$ and $w_1$ are consistent with $\Lambda\textrm{CDM}$
whereas $1\sigma$ limits of $w_2$ deviate from $w=-1$. In order to
estimate the statistical significance of the trends, we have also
analyzed $\Lambda\textrm{CDM}$ based Constitution mock data sets,
and estimated $68.3$\% and $95.4$\% confidence limits on $w_i$'s
based on the total sum of five thousand MCMC chains; these are shown
as grey regions in Fig.\ \ref{fig:2trz-constitution-mock} (top panels).
The likelihood confidence regions for the $\Lambda\textrm{CDM}$ mock
data sets are wider than those determined from the marginalized
one-dimensional likelihood distributions for the Constitution data
set (red error bars). This is because the mock data sets include the
additional statistical variance caused by different realizations of
measurement errors.

As a second measure of average, we compute the arithmetic mean
(denoted as M) of the MCMC chain elements for each cosmological parameter.
In fact, it is known that in the MCMC method the arithmetic mean of MCMC
chain is appropriate as the best estimate for the parameter and is a more
robust quantity than the peak (or mode) of the marginalized distribution
\cite{Spergel-etal-2003}.
The bottom row of Fig.\ \ref{fig:2trz-constitution-mock} shows distributions
of five thousand M values of $w_0$, $w_1$, and $w_2$, each estimated from
the MCMC chain of individual mock data set. The smooth distributions
(shown as dark red curves) have been generated by applying the Savitzky-Golay
smoothing filter \cite{numerical-recipe} to the histograms with the bin size
$\Delta w=0.02$.
The distribution for $w_1$ is wider than those for $w_0$ and $w_2$,
implying that the equation of state parameter $w_1$ is less tightly
constrained in our double-transition model with $z_{\textrm{tr}1}=0.2$;
this is due to different number of data in each bin.
For a comparison, M values for the observed SNIa data sets are indicated
by vertical lines.
In Constitution and Constitution-U sets all M values of $w_i$'s are
outside $1 \sigma$ regions of the distributions. In the Union set,
the M values of $w_0$ and $w_1$ are consistent with the $\Lambda$CDM mock
data, whereas that of $w_2$ is outside the $1 \sigma$ region. Although
we detect more than $1\sigma$ deviations of all $w_i$'s in the Constitution
and the Constitution-U sets, we would like to emphasize that all the
deviations are still within $2 \sigma$ regions, consistent with
$\Lambda\textrm{CDM}$ model. The similarity between Constitution and
Constitution-U sets and the difference between Constitution-U and Union
are important because the Constitution-U and the Union sets are based
on {\it the same} SALT light-curve fit parameters of the Union SNIa sample.
The different behaviors of $w_0$ and $w_1$ between the Constitution-U
and the Union sets are purely due to different calibration processes
applied in producing the SNIa distance modulus (see Sec.\ \ref{subsec:obsdata}).

\begin{table*}
\begin{center}
\caption{
         Estimated $w_0$, $w_1$, $w_2$, and $\Omega_{\textrm{m}0}$ in
         the double-transition dark energy model for $z_{\rm tr1} = 0.2$
         and $z_{\rm tr2}=0.4$. For comparison, results from the five thousand
         $\Lambda\textrm{CDM}$ based Constitution mock data sets are
         presented in the bottom.
         Symbols in the parenthesis represent our methods of average
         estimation, and the errors indicate $68.3$\% confidence limits
         (see text).
         }
\begin{tabular} {cccccc}
 \\
 \hline\\[-3.0mm]
 \quad Data     &   Average     &     \quad $w_0$         &      \quad  $w_1$       &      \quad  $w_2$       &    $\Omega_{\textrm{m}0}$    \\[0.5mm] \hline\\[-2.5mm]
 Constitution   &  (P)  & $-0.71_{-0.16}^{+0.15}$ & $-1.78_{-0.43}^{+0.42}$ & $-0.54_{-0.23}^{+0.18}$ &   $0.292_{-0.018}^{+0.017}$  \\[0.5mm]
 Constitution-U &  (P)  & $-0.73_{-0.20}^{+0.19}$ & $-1.73_{-0.45}^{+0.45}$ & $-0.55_{-0.23}^{+0.19}$ &   $0.291_{-0.019}^{+0.019}$  \\[0.5mm]
 Union          &  (P)  & $-0.98_{-0.22}^{+0.20}$ & $-1.11_{-0.54}^{+0.42}$ & $-0.63_{-0.27}^{+0.19}$ &   $0.283_{-0.018}^{+0.017}$  \\[0.5mm]
 Union2         &  (P)  & $-0.97_{-0.12}^{+0.11}$ & $-1.22_{-0.33}^{+0.30}$ & $-0.71_{-0.25}^{+0.21}$ &   $0.278_{-0.017}^{+0.014}$  \\[0.5mm]
 \hline\\[-3.0mm]
 Constitution   & (M) & $-0.71_{-0.16}^{+0.15}$ & $-1.79_{-0.42}^{+0.43}$ & $-0.62_{-0.15}^{+0.26}$ &   $0.294_{-0.016}^{+0.015}$   \\[0.5mm]
 Constitution-U & (M) & $-0.74_{-0.19}^{+0.18}$ & $-1.74_{-0.44}^{+0.46}$ & $-0.64_{-0.13}^{+0.28}$ &   $0.292_{-0.018}^{+0.017}$   \\[0.5mm]
 Union          & (M) & $-1.00_{-0.20}^{+0.22}$ & $-1.18_{-0.47}^{+0.49}$ & $-0.73_{-0.17}^{+0.29}$ &   $0.285_{-0.020}^{+0.015}$   \\[0.5mm]
 Union2         & (M) & $-0.97_{-0.12}^{+0.11}$ & $-1.23_{-0.32}^{+0.31}$ & $-0.79_{-0.17}^{+0.29}$ &   $0.279_{-0.018}^{+0.013}$   \\[1.0mm]
 \hline\\[-3.0mm]
 Constitution mock  &  (PP)  & $-1.02_{-0.17}^{+0.17}$  &   $-0.93_{-0.47}^{+0.43}$  &   $-0.92_{-0.35}^{+0.30}$  &   $0.277_{-0.017}^{+0.015}$  \\[0.5mm]
 Constitution mock  &  (PM)  & $-1.03_{-0.17}^{+0.17}$  &   $-0.92_{-0.45}^{+0.46}$  &   $-1.02_{-0.47}^{+0.28}$  &   $0.278_{-0.017}^{+0.014}$   \\[0.5mm]
 Constitution mock  &  (MM)  & $-1.03_{-0.17}^{+0.17}$  &   $-0.91_{-0.46}^{+0.45}$  &   $-1.23_{-0.26}^{+0.49}$  &   $0.278_{-0.017}^{+0.014}$   \\[0.5mm]
 \hline
\end{tabular}
\label{Table:double-trz-0204}
\end{center}
\end{table*}

Table \ref{Table:double-trz-0204} compares equation of state
parameters and matter density parameter in the double-transition
dark energy model with $z_{\textrm{tr}1}=0.2$ and
$z_{\textrm{tr}2}=0.4$ for various SNIa data sets and average
estimation methods. For $\Lambda\textrm{CDM}$ based Constitution
mock data sets, first we obtain distributions of P or M values and
then present the peak of P distribution (PP), the peak of M
distribution (PM), and the mean of M distribution (MM), together
with $68.3$\% confidence limits. The location of peak and the
confidence limits are determined based on the smooth distribution
generated by applying the Savitzky-Golay smoothing filter to the
histograms of P or M values. Comparison of the parameters estimated
by three different average methods (PP, PM, and MM) with the fiducial
value $w=-1$ suggests that $w_1$ is biased to the positive direction
by the amount of about $0.07$--$0.09$. The estimated bias is
uncertain roughly within $\sigma/\sqrt{N} \simeq
0.45/\sqrt{5000}=0.006$. The $w_0$ is very weakly biased to the
negative direction. For $w_2$ parameter, the presence of bias is not
clear at this point because the three average estimates PP, PM, and
MM indicate positive, zero-consistent, and negative biases relative
to $w=-1$, respectively.

We notice all the observational data sets show a trend that P value
is larger than M value for $w_2$ parameter, $w_2~(\textrm{P})
> w_2~(\textrm{M})$ unlike other parameters.
The similar feature is seen in the results of Constitution mock data
analysis (PP and PM values in Table \ref{Table:double-trz-0204}). As
shown in bottom row of Fig.\ \ref{fig:2trz-constitution-mock}, in
the case of Constitution mock data sets M distributions for $w_0$
and $w_1$ look symmetric while that for $w_2$ has a non-symmetric
shape with a tail in the direction to which $w_2$ decreases. We
expect from the PP value together with $\pm1\sigma$ errors listed in
Table \ref{Table:double-trz-0204} that the P distribution for $w_2$
is also asymmetric and has a tail in the same direction and a peak
location larger than that of the M distribution. In the increasing
direction of $w_2$ the parameter is constrained by the big-bang
nucleosynthesis calculation and the CMB shift parameter. On the
other hand, $w_2$ is free to randomly walk in the opposite direction
without a strong prior like $w_2 > -5$ (see Sec.\
\ref{sec:method-data}) since the subdominant dark energy component
at high redshift has more freedom to take any value of equation of
state parameter. That is the reason why the likelihood distribution
of $w_2$ parameter has a tendency of having the asymmetric shape
with a tail to the smaller $w_2$ direction. Thus, it is naturally
expected that the location of peak (P) in the one-dimensional
marginalized distribution is larger than the arithmetic mean (M) of
MCMC chain elements.

\begin{figure*}
\begin{center}
\includegraphics[width=14cm]{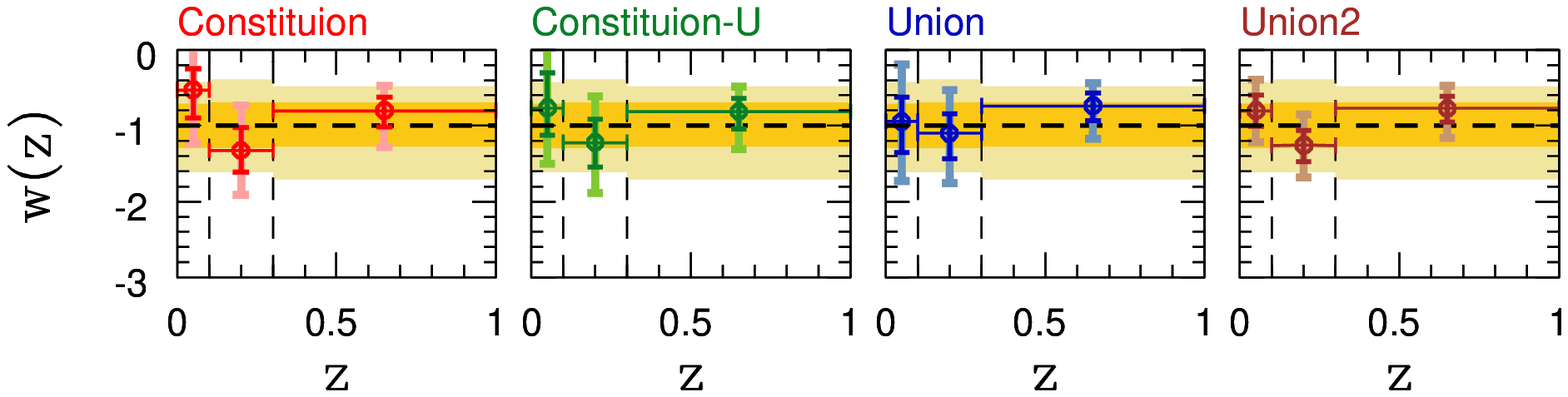}
\includegraphics[width=14cm]{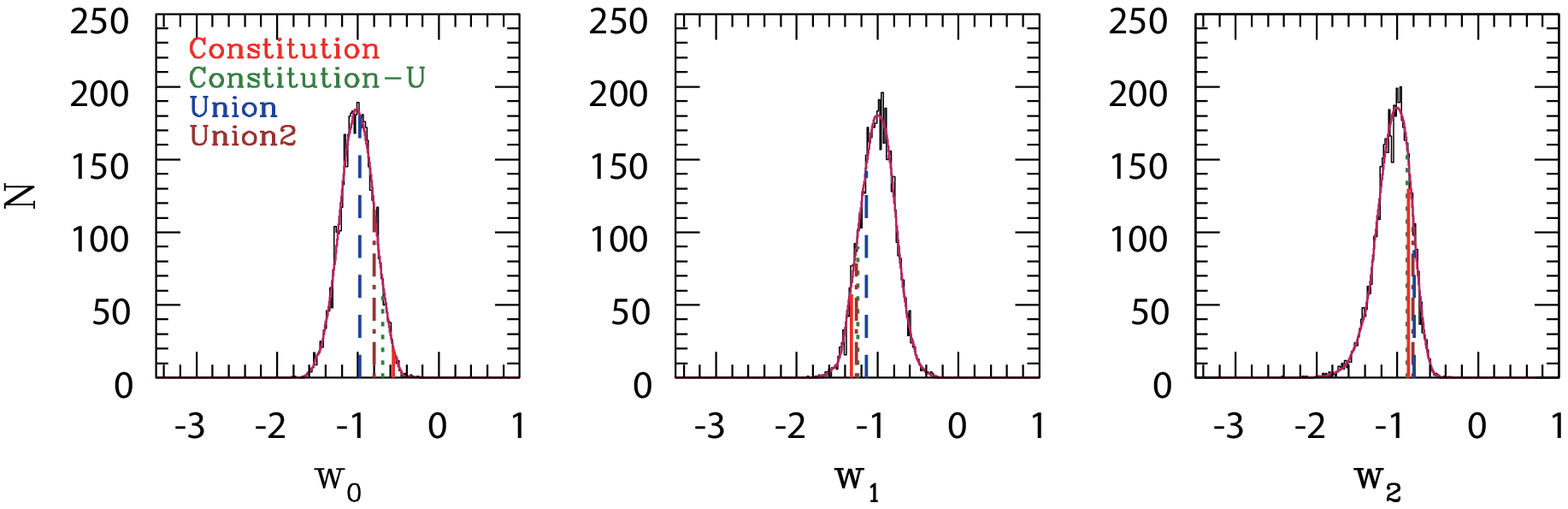}
\caption{
         The same as Fig.\ \ref{fig:2trz-constitution-mock} but for
         $z_{\textrm{tr}1}=0.1$ and $z_{\textrm{tr}2}=0.3$ and
         based on Union2 mock data sets.
         The three bins in Union2 set have $165$, $117$, and $275$ SNIa
         data points, respectively.
         In the top panels, data points with
         error bars are based on the actual SNIa data sets while
         the shaded regions (in all panels) are obtained from the five
         thousand $\Lambda\textrm{CDM}$ based Union2 mock data sets.
         }
\label{fig:2trz-union2-mock}
\end{center}
\end{figure*}

\begin{table*}
\begin{center}
\caption{The same as Table \ref{Table:double-trz-0204} but
        for $z_{\textrm{tr}1}=0.1$ and $z_{\textrm{tr}2}=0.3$
        and based on Union2 mock data sets.}
\begin{tabular} {cccccc}
 \\
 \hline\\[-3.0mm]
 \quad Data     &  Average   &        \quad $w_0$        &         \quad $w_1$        &       \quad $w_2$          &    $\Omega_{\textrm{m}0}$    \\[0.5mm]
 \hline\\[-2.5mm]
 Constitution   &  (P)  &  $-0.53_{-0.37}^{+0.28}$  &   $-1.33_{-0.29}^{+0.30}$  &   $-0.80_{-0.21}^{+0.18}$  &   $0.285_{-0.017}^{+0.017}$  \\[0.5mm]
 Constitution-U &  (P)  &  $-0.77_{-0.36}^{+0.46}$  &   $-1.22_{-0.32}^{+0.31}$  &   $-0.82_{-0.22}^{+0.18}$  &   $0.280_{-0.017}^{+0.020}$  \\[0.5mm]
 Union          &  (P)  &  $-0.95_{-0.41}^{+0.32}$  &   $-1.10_{-0.33}^{+0.25}$  &   $-0.74_{-0.20}^{+0.17}$  &   $0.280_{-0.017}^{+0.019}$  \\[0.5mm]
 Union2         &  (P)  &  $-0.80_{-0.20}^{+0.21}$  &   $-1.26_{-0.21}^{+0.20}$  &   $-0.77_{-0.18}^{+0.16}$  &   $0.278_{-0.015}^{+0.015}$  \\[0.5mm]
 \hline\\[-2.5mm]
 Constitution   & (M) &  $-0.56_{-0.34}^{+0.31}$  &   $-1.32_{-0.30}^{+0.29}$  &   $-0.86_{-0.15}^{+0.26}$  &   $0.286_{-0.016}^{+0.018}$  \\[0.5mm]
 Constitution-U & (M) &  $-0.70_{-0.43}^{+0.39}$  &   $-1.24_{-0.30}^{+0.33}$  &   $-0.88_{-0.16}^{+0.24}$  &   $0.283_{-0.014}^{+0.023}$  \\[0.5mm]
 Union          & (M) &  $-0.98_{-0.38}^{+0.35}$  &   $-1.14_{-0.29}^{+0.29}$  &   $-0.79_{-0.15}^{+0.22}$  &   $0.282_{-0.019}^{+0.017}$  \\[0.5mm]
 Union2         & (M) &  $-0.80_{-0.20}^{+0.21}$  &   $-1.27_{-0.20}^{+0.21}$  &   $-0.81_{-0.14}^{+0.20}$  &   $0.280_{-0.017}^{+0.013}$  \\[1.0mm]
 \hline\\[-2.5mm]
 Union2 mock    &  (PP)  &  $-1.00_{-0.21}^{+0.21}$  &   $-1.00_{-0.22}^{+0.21}$  &   $-0.95_{-0.22}^{+0.18}$  &   $0.276_{-0.014}^{+0.015}$  \\[0.5mm]
 Union2 mock    &  (PM)  &  $-1.00_{-0.22}^{+0.21}$  &   $-1.00_{-0.22}^{+0.22}$  &   $-1.01_{-0.22}^{+0.20}$  &  $0.278_{-0.016}^{+0.015}$     \\[0.5mm]
 Union2 mock    &  (MM)  &  $-1.01_{-0.21}^{+0.20}$  &   $-1.00_{-0.22}^{+0.22}$  &   $-1.06_{-0.17}^{+0.25}$  &  $0.278_{-0.016}^{+0.015}$     \\[0.5mm]
 \hline
\end{tabular}
\label{Table:double-trz-0103}
\end{center}
\end{table*}

We made similar analysis based on Union2 mock data sets, and the
results are shown in Fig.\ \ref{fig:2trz-union2-mock} and Table
\ref{Table:double-trz-0103}. Here, we choose the transition
redshifts $z_{\textrm{tr}1}=0.1$ and $z_{\textrm{tr}2}=0.3$, for
which the deviations of $w_i$'s from $w=-1$ are the largest in the
case of Union2 data set (see Fig.\ \ref{fig:2trz-data-plot}). In the
case of Union2 set, $1\sigma$ confidence limits of $w_1$ and $w_2$
deviate from $w=-1$ whereas $w_0$ is consistent with
$\Lambda\textrm{CDM}$. However, all the values of $w_i$'s from the
Union2, Union, Constitution, and Constitution-U sets are within
$2\sigma$ limits thus are consistent with $\Lambda\textrm{CDM}$
model (top row of Fig.\ \ref{fig:2trz-union2-mock}). We also see
that compared with the M distributions for $\Lambda$CDM based Union2
mock data sets all M values of $w_i$'s are consistent with
$\Lambda$CDM model and deviations are not statistically significant.
The M distributions of $w_1$ and $w_2$ (bottom row of Fig.\
\ref{fig:2trz-union2-mock}) are narrower than those for Constitution
mock data sets (bottom row of Fig.\
\ref{fig:2trz-constitution-mock}), reflecting that the dark energy
equation of state parameters are relatively well constrained as we
have more SNIa data points in the redshift bins.

We observe the same trend that P value is larger than M value for $w_2$
parameter, $w_2~(\textrm{P}) > w_2~(\textrm{M})$.
But the difference between P and M values is smaller due to the increase
of the width of the final redshift bin ($z \ge 0.3$) and of SNIa data
points within the bin.
As shown in Table \ref{Table:double-trz-0103}, the Union2 mock data analysis
implies that estimated $w_0$ and $w_1$ are unbiased by three average
methods (PP, PM, and MM); they are almost the same as the fiducial value $w=-1$.
The PP, PM, and MM average methods for $w_2$ parameter give different
values, indicating the likelihood distribution for this parameter is also not
symmetric. However, the asymmetry is weaker than the case of Constitution
data set.

\section{Characters of current SNIa data based on mock data analysis}
\label{sec:charac}

In the double-transition model the Constitution mock data sets shows
some statistical deviation of equation of state parameters from
the fiducial $\Lambda$CDM model (Table \ref{Table:double-trz-0204}).
For PM average, the value $w_1=-0.92$ indicates a bias at intermediate
redshift interval although it is small compared with the size of $1\sigma$
error. In the Constitution sample, the SNIa data points are irregularly
distributed between $z=0.015$ and $z=1.551$ with non-uniform distance
modulus errors, and the data is sparse in the intermediate redshift
$z=0.1$--$0.3$.
In this section, we investigate a bias that may be induced by the
characters of the current SNIa data sets. The estimation of the bias
is important because it can be misinterpreted as the spurious
evolution of dark energy.

We have generated $\Lambda\textrm{CDM}$ based mock data sets in three
different ways. First, we made SNIa data with uniform distribution in
the redshift interval and uniform distance modulus errors (Case 1).
We arrange SNIa data points ($N_\textrm{SN}=397$) uniformly on the redshift
range spanned by the Constitution sample and assign the uniform error
to all members. We take the harmonic average as the uniform error given by
\begin{equation}
    \overline{\sigma}
       =\sqrt{\frac{N_\textrm{SN}}{\sum_{i=1}^{N_\textrm{SN}}
          1/\sigma_{\textrm{obs},i}^{2}}}=0.177,
\label{eq:error-average}
\end{equation}
in magnitude scale. The second-type mock data sets have the uniform
redshift-distribution but with distance modulus errors of the
Constitution sample (Case 2). We sort the Constitution SNIa data
points in increasing order in redshift, then assign distance modulus
errors to mock SNIa members in the same order. The third-type mock
data sets have the same redshift-distribution as the Constitution
data set but with uniform distance modulus errors (Case 3). For
comparison, we include the Constitution mock data sets as Case 4;
these are already presented in Table \ref{Table:double-trz-0204}. We
have analyzed these mock data sets (including mock BAO and CMB
parameters) in the same way as the Constitution mock data sets have
been analyzed for double-transition model with
$z_{\textrm{tr}1}=0.2$ and $z_{\textrm{tr}2}=0.4$ (Sec.\
\ref{sec:double}). The results are shown in Table
\ref{Table:compare-mock}, which compares dark energy equation of
state parameters obtained by using different average estimation
methods (PP, PM, and MM).

\begin{table*}
\begin{center}
\caption{Dark energy equation of state parameters of double-transition
        model with $z_{\rm tr1} = 0.2$ and $z_{\rm tr2}=0.4$ estimated
        from the five thousand $\Lambda\textrm{CDM}$ based SNIa mock data
        sets with four different types (Cases 1--4; see text).
        Values are listed based on three average estimation methods
        (PP, PM, and MM). The errors indicate $68.3$\% confidence levels.}
\begin{tabular} {lccc}
\\
\hline\\[-3.0mm]
 Data                           &      \quad $w_0$ (PP)       &      \quad $w_1$ (PP)       &     \quad $w_2$ (PP)     \\[0.5mm] \hline\\[-2.5mm]
 Case 1: uniform distribution and error &   $-0.99_{-0.27}^{+0.27}$  &   $-1.00_{-0.47}^{+0.45}$  &   $-0.96_{-0.25}^{+0.22}$   \\[0.5mm]
 Case 2: uniform distribution           &   $-0.97_{-0.30}^{+0.29}$  &   $-1.03_{-0.47}^{+0.46}$  &   $-0.95_{-0.25}^{+0.22}$   \\[0.5mm]
 Case 3: uniform error                  &   $-1.01_{-0.19}^{+0.18}$  &   $-0.97_{-0.50}^{+0.46}$  &   $-0.92_{-0.34}^{+0.27}$   \\[0.5mm]
 Case 4: Constitution mock              &   $-1.02_{-0.17}^{+0.17}$  &   $-0.93_{-0.47}^{+0.43}$  &   $-0.92_{-0.35}^{+0.30}$   \\[1.0mm]
 \hline\\[-3.0mm]
 Data                           &      \quad $w_0$ (PM)      &       \quad $w_1$ (PM)      &      \quad $w_2$ (PM)     \\[0.5mm] \hline\\[-2.5mm]
 Case 1: uniform distribution and error &   $-1.00_{-0.28}^{+0.28}$  &   $-1.00_{-0.47}^{+0.48}$  &   $-1.01_{-0.29}^{+0.24}$     \\[0.5mm]
 Case 2: uniform distribution           &   $-0.98_{-0.31}^{+0.30}$  &   $-1.01_{-0.49}^{+0.45}$  &   $-1.03_{-0.27}^{+0.25}$     \\[0.5mm]
 Case 3: uniform error                  &   $-1.01_{-0.19}^{+0.17}$  &   $-0.95_{-0.49}^{+0.48}$  &   $-1.02_{-0.27}^{+0.24}$     \\[0.5mm]
 Case 4: Constitution mock              &   $-1.03_{-0.17}^{+0.17}$  &   $-0.92_{-0.45}^{+0.46}$  &   $-1.02_{-0.47}^{+0.28}$     \\[0.5mm]
 \hline\\[-3.0mm]
 Data                           &      \quad $w_0$ (MM)      &       \quad $w_1$ (MM)      &      \quad $w_2$ (MM)     \\[0.5mm] \hline\\[-2.5mm]
 Case 1: uniform distribution and error &   $-1.00_{-0.28}^{+0.28}$  &   $-1.01_{-0.46}^{+0.49}$  &   $-1.10_{-0.20}^{+0.33}$     \\[0.5mm]
 Case 2: uniform distribution           &   $-0.98_{-0.31}^{+0.30}$  &   $-1.04_{-0.46}^{+0.48}$  &   $-1.08_{-0.22}^{+0.30}$     \\[0.5mm]
 Case 3: uniform error                  &   $-1.03_{-0.17}^{+0.19}$  &   $-0.94_{-0.48}^{+0.47}$  &   $-1.17_{-0.12}^{+0.39}$     \\[0.5mm]
 Case 4: Constitution mock              &   $-1.03_{-0.17}^{+0.17}$  &   $-0.91_{-0.46}^{+0.45}$  &   $-1.23_{-0.26}^{+0.07}$     \\[0.5mm]
 \hline
\end{tabular}
\label{Table:compare-mock}
\end{center}
\end{table*}

\begin{figure*}
\begin{center}
\includegraphics[width=14cm]{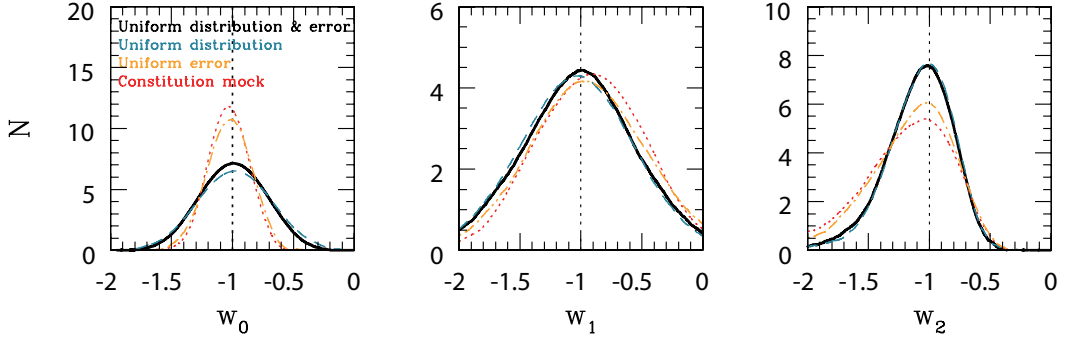}
\caption{
   Distributions of M values for four different types of
   $\Lambda\textrm{CDM}$ mock data sets. The smooth distributions have been
   made based on the histograms with the bin size $\Delta w = 0.001$.
   Black (solid), blue (dashed), yellow (dot-dashed), and red (dotted) curves represent
   Case 1 (uniform distribution \& error), Case 2 (uniform distribution),
   Case 3 (uniform error), and Case 4 (Constitution mock), respectively.
   Vertical dashed lines indicate the location of $w=-1$.}
\label{fig:compare-mock}
\end{center}
\end{figure*}

For all Cases 1--4, we see that $w_0$ and $w_1$ almost do not depend
on the choice of average methods, that is,
$w_i~(\textrm{PP}) \simeq w_i~(\textrm{PM}) \simeq w_i~(\textrm{MM})$,
while $w_2$ parameters show a trend shown in the previous section,
$w_2~(\textrm{PP}) > w_2~(\textrm{PM}) > w_2~(\textrm{MM})$.
The likelihood distributions of $w_0$ and $w_1$ are symmetric whereas
that of $w_2$ is not symmetric for all cases of mock data sets. This
is verified in Fig.\ \ref{fig:compare-mock}, which compares distributions
of M values derived from the four cases of $\Lambda\textrm{CDM}$
mock data sets.
Expecting that analysis of Case 1 (uniform distribution and error) restores
the fiducial value $w=-1$ correctly, we found that the PM acts as
the unbiased (or less biased) estimator for the average of parameter
because PM values of $w_0$, $w_1$, and $w_2$ are consistent with the
$\Lambda$CDM model. Even for the non-symmetric shape of likelihood
distribution as in the case of $w_2$ parameter, the PM average is
unbiased, and thus provides the precise estimation of bias in the parameter.

For Constitution mock data sets (Case 4), the bias in the positive direction
for $w_1$ parameter is observed. Because the mock data sets of Case 3
(uniform error) induce the similar amount of bias in the same direction while
the data sets of Case 2 (uniform distribution) does not affect $w_1$ much,
we conclude that the irregular distribution of Constitution SNIa data points
is more concerned with the bias than the non-uniform distance modulus errors.
However, the bias estimated based on the $\Lambda\textrm{CDM}$ mock data
sets is small compared with the precision the current SNIa data (including
BAO and CMB parameters) allows. For Constitution mock data (Case 4),
it amounts to $1+w_1\simeq 0.18\sigma$ at most for PM average.
Unlike the bias in the $w_1$ parameter, $w_2$ is not much biased (based
on PM average).

For the Union2 mock data sets, we have not seen any significant
deviation of equation of state parameters (see Table
\ref{Table:double-trz-0103}). Because of the relatively denser SNIa data
points, the biases in $w_i$ estimation are smaller than those of
Constitution case. However, the likelihood distribution of $w_2$
still appears asymmetric (as shown in Fig.\
\ref{fig:2trz-union2-mock}), and thus the estimated value of $w_2$
depends on the average methods.

%
%
\section{Discussion}
\label{sec:discussion}

In this study, we have investigated the evolution of dark energy
equation of state parameter $w(z)$ using the sudden-jump
approximation of $w$ for some chosen redshift intervals with double
transitions. We used four SNIa data (Constitution, Constitution-U,
Union, and Union2) together with BAO $A$ parameter and CMB $R$
parameter, and used the MCMC method with Metropolis-Hastings
algorithm to obtain the likelihood in the parameter space. In order
to test the statistical significance of, or to estimate the bias in,
the equation of state parameters we investigated the case of
$\Lambda$CDM mock data sets. Since the analysis of thousands of
$\Lambda\textrm{CDM}$ mock data sets can include all statistical
effects due to the characters of data sets such as the irregular
distribution in the redshift interval and the different random
realization of measurement errors, our method of using the
$\Lambda\textrm{CDM}$ mock data is complementary to the uncorrelated
estimation of dark energy equation of state parameters based on the
decorrelating technique \cite{Huterer-etal-2002,Huterer-etal-2004,Sullivan-etal-2007}.

In the double-transition dark energy models, we show that
there are some deviations from $\Lambda\textrm{CDM}$ in the low
($z \lesssim 0.2$) and the middle
($0.2 \lesssim z \lesssim 0.4$) redshift regions especially in the
Constitution and Constitution-U sets.
Deviations of $w_i$'s from $\Lambda$CDM in the low
and the middle redshift are between $1$ and $2 \sigma$ confidence
levels, see the top row of Fig.\ \ref{fig:2trz-constitution-mock}.
The deviations are stronger when we compare each arithmetic mean (M) value
of $w_i$ with the $\Lambda$CDM mock data;
the M values of Constitution and Constitution-U sets are situated near
$2\sigma$ boundary of the $\Lambda$CDM mock data distribution, see
the bottom row of Fig.\ \ref{fig:2trz-constitution-mock}, whereas
there are no such deviations in the Union set. Thus, we conclude that
our analysis of the double-transition model indicates statistically
noticeable deviation of the Constitution data from $\Lambda$CDM model
in the low redshift. Furthermore, consistent result between
Constitution set and Constitution-U set suggests that the detected
deviation in the Constitution data is not due to additional data in the
Constitution data, but mainly due to different calibration of the same Union
data.
Results of Constitution-U compared with the Constitution and the Union data
in Figs.\ \ref{fig:2trz-data-plot} and \ref{fig:2trz-constitution-mock}
apparently show that the new 90 CfA3 SNIa data added in the
Constitution data do not have role in causing the difference between
the Constitution and the Union sets.

\begin{figure*}[htbp]
\begin{center}
\includegraphics[width=14cm]{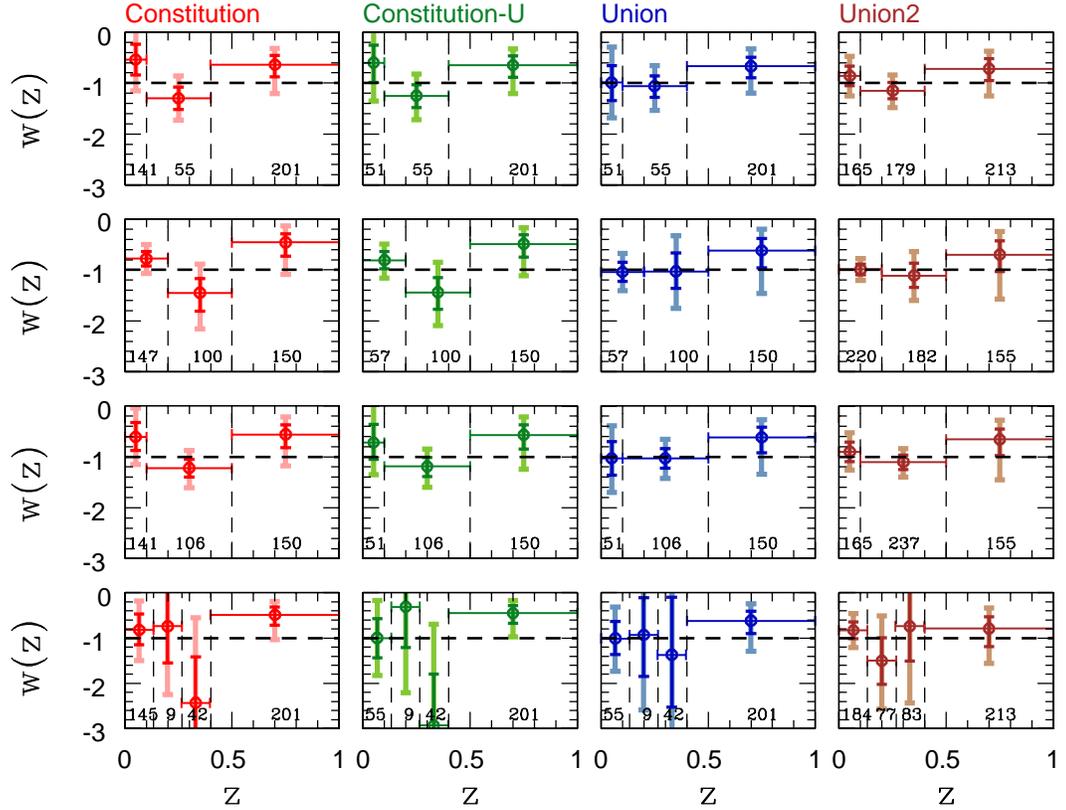}
\caption{Top three rows:
         dark energy equation of state parameters ($w_0$, $w_1$, $w_2$)
         measured from Constitution, Constitution-U, Union, and Union2 samples
         for different choices of transition redshifts, $z_\textrm{tr1}$ and
         $z_\textrm{tr2}$. In order from top to bottom we choose
         ($a$) $z_\textrm{tr1}=0.1$ and $z_\textrm{tr2}=0.4$,
         ($b$) $z_\textrm{tr1}=0.2$ and $z_\textrm{tr2}=0.5$, and
         ($c$) $z_\textrm{tr1}=0.1$ and $z_\textrm{tr2}=0.5$.
         Bottom row: ($d$) dark energy equation of state parameters
         in the triple-transition dark energy model. The first three bins
         are equally spaced in the redshift interval $z=0$--$0.4$.
         Each number listed in each redshift bin indicates the number of
         SNIa included in the bin. Note that although the size of error bars
         generally gets smaller as the corresponding redshift bin includes
         more SNIa data points, it is also affected by the quality of
         SNIa data points within the bin.
         }
\label{fig:2-3trz}
\end{center}
\end{figure*}

Our conclusion is still valid when we measure dark energy equation of state
parameters by choosing different transition redshifts, as summarized
in Fig.\ \ref{fig:2-3trz} (top three rows), where the width of the second
redshift bin has been extended by the amount of $\Delta z=0.1$ to left,
right, and both directions.
Although extending the redshift interval of the middle redshift bin makes
the error bar of $w_1$ value smaller as expected,
the observed trend for $w_0$ and $w_1$ estimates of Constitution,
Constitution-U, and Union samples does not change much in comparison with
the result shown in Fig.\ \ref{fig:2trz-constitution-mock}.

In the bottom row of Fig.\ \ref{fig:2-3trz} we present our measurement
of equation of state parameters based on the triple-transition model with
the first three redshift bins equally spaced between $z=0$ and $0.4$.
Even in this case our conclusion maintains. However, increasing number
of redshift bins causes error bars to be drastically larger and it is
more and more difficult to make the MCMC chains converge without strong
priors. This is why we have chosen a simple dark energy model with double
transitions.

We emphasize that the Union and the Constitution-U sets are composed
of the same SNIa members and are based on the same light-curve fit parameters.
In a dark energy model with constant equation of state,
the $w$ parameter (P value) estimated from SNIa data together with
BAO $A$ parameter is $1+w=-0.011_{-0.080}^{+0.074}$ for Union sample and
$1+w=-0.010_{-0.080}^{+0.078}$ for Constitution-U sample, both of which
are nearly identical to each other and are similar to the values
obtained by Refs.\ \cite{Kowalski-etal-2008,Hicken-etal-2009}.
Although the two data sets give nearly identical results in the simple
constant $w$ dark energy model, they show different behaviors of
dark energy equation of state in our evolving dark energy model with double
transitions, due to different calibrations applied during the production
of SNIa distance modulus.
Our analysis suggests that it is generally important
to distinguish the real $w$ evolution from the calibration artefact in the
SNIa data analysis.

However, we cannot conclude that such deviations in the Constitution
data are the strong hint for evolution of dark energy because all the
equation of state parameters $w_i$'s measured from all the observational
data sets are consistent with $\Lambda\textrm{CDM}$
model within $2\sigma$ confidence limits.
Our results are also consistent with previous studies searching for
dark energy evolution with SNIa data based on the redshift-binning method
\cite{Serra-etal-2009,Huang-etal-2009,Gong-etal-2009,Cai-etal-2010,
Zhao-etal-2009,Amanullah-etal-2010,Wang-etal-2010,Qi-etal-2009}.
Furthermore, the difference between Union and Constitution-U due to
different calibrations is also not significant at the level of current
precision of SNIa data.
Such an effect caused by different calibrations should be considered to
be important in the forthcoming missions of SNIa survey.

From the analysis of $\Lambda\textrm{CDM}$-motivated mock data sets
with four different types (Cases 1--4), we have tried to estimate
bias in the equation of state parameters due to the character of
data set used in search for dark energy evolution. Comparing average
values for $w_i$'s obtained by three different methods such as PP
(peak of peaks), PM (peak of means), and MM (mean of means) while
knowing the answer ($w=-1$), we conclude that the PM is an unbiased
estimator of average bias. For the symmetric likelihood
distribution, all average methods give values consistent with each
other (e.g., see Fig.\ \ref{fig:2trz-constitution-mock} and $w_0$ and $w_1$
for the Constitution data sets in Table \ref{Table:compare-mock}).
However, for the non-symmetric likelihood distribution as in the
case of $w_2$ parameter, the PP and MM averages are biased to the
opposite directions with respect to the PM average which restores
the fiducial value $w=-1$ correctly. The Constitution sample turns
out to induce a bias in the $w_1$ parameter by the amount of about
$0.2\sigma$ whereas the Union2 sample does not show any noticeable
bias in the estimation of parameters (see Table
\ref{Table:double-trz-0103}). In order to precisely estimate any
bias due to the character of data set based on the mock data
analysis, we suggest to use the PM quantity, the peak (or mode) of
the distribution of arithmetic means, as the unbiased estimator for
the average bias.

%
%
\acknowledgments
We wish to thank Dr. Eric V. Linder, Dr. Yong-Seon Song,
and Professor Dal Ho Kim for useful comments and discussions.
We also would like to thank the anonymous referee
for the constructive and helpful comments on our manuscript.
This work was supported by the Korea Research Foundation
Grant funded by the Korean Government
(KRF-2008-341-C00022).

%
%


\end{document}